\documentclass[a4paper,10pt,twocolumn]{article}
\usepackage{flushend}
\usepackage[left=1.8cm, right=1.5cm, top=2cm, bottom=2.5cm]{geometry}
\usepackage[T1]{fontenc}
\usepackage{lmodern}
\usepackage{amssymb,amsmath}
\usepackage[latin1]{inputenc}
\usepackage[english]{babel}
\usepackage{graphicx}
\usepackage{color}
\usepackage[super, comma, sort&compress]{natbib}
\usepackage{hyperref}
\hypersetup{
  pdfauthor={Stefan Kremer, Raymond Fr\'esard},
  pdftitle={Thermoelectric transport properties of an apparent Fermi liquid: Relation to an analytic anomaly in the density of states and application to hole-doped delafossites},
  pdfsubject={doi:10.1002/andp.201100165},
  pdfkeywords={low-dimensional compounds, thermoelectricity, electronic structure, delafossites},
  pdfproducer={Copyright \textcopyright ~(2011) Stefan Kremer and Raymond Fr\'esard. All rights reserved.},
  pdfborder={0 0 0}
}

\renewcommand{\thesection}{\Roman{section}}
\makeatletter
\renewcommand{\section}{
	\@ifstar\sectionStar\sectionNoStar
}
\makeatother
\newcommand{\sectionStar}[2]{
 	\par
	\vspace{1em}
	\pagebreak[3]
	\begin{center}
		\textbf{\MakeUppercase{#1}}
		\addcontentsline{toc}{section}{#1}
	\end{center}
	\vspace{1em}
}
\newcommand{\sectionNoStar}[1]{
	\par
	\vspace{0.5em}
	\pagebreak[3]
	\refstepcounter{section}
	\begin{center}
		\textbf{\thesection .~\MakeUppercase{#1}}
		\addcontentsline{toc}{section}{#1}
	\end{center}
	\vspace{0.5em}
}
\renewcommand{\thesubsection}{\Roman{section}.\Alph{subsection}}
\renewcommand{\subsection}[1]{
	\par
	\vspace{5pt}
	\refstepcounter{subsection}
	\begin{center}
		\normalsize\textbf{\thesubsection .~{#1}}
		\addcontentsline{toc}{subsection}{#1}
	\end{center}
	\vspace{5pt}
}


\begin{document}

	\title{~\\[-3em]
		\begin{picture}(0,0)(0,0)
			\setlength{\unitlength}{\textheight}
			\put(0,-1){\makebox(0,0)[c]{\tiny Copyright \textcopyright ~(2011) Stefan Kremer and Raymond Fr\'esard. All rights reserved.}}
		\end{picture}~\\[-0.9em]
		\large \textbf{Thermoelectric transport properties of an apparent Fermi liquid:
		Relation to an analytic anomaly in the density of states and application to hole-doped delafossites}
	}

	\author{Stefan Kremer\textsuperscript{1,2,\!\!}
		\footnote{Corresponding author\quad
			E-mail:~\textsf{stefan.kremer@kit.edu}}
		~and Raymond Fr\'esard\textsuperscript{2}\\[0.5em]
		\textit{\normalsize\textsuperscript{1}Institut f\"ur Theorie der Kondensierten Materie, Karlsruhe Institute of Technology (KIT),}\\
		\textit{\normalsize 76128 Karlsruhe, Germany}\\
		\textit{\normalsize\textsuperscript{2}Laboratoire CRISMAT, UMR CNRS-ENSICAEN (ISMRA) 6508, 14050 Caen, France}\\[0.5em]
		{\normalsize (Published in Ann. Phys. (2011), doi:10.1002/andp.201100165)}\\
	}

	\date{\normalsize
	\parbox{0.85\textwidth}{
		Through the motivation of the recent discovery of dispersionless regions in the band structure of the delafossites, a model density of states of free fermions including a discontinuity as analytical anomaly is studied.
		The resulting temperature dependence of the chemical potential is obtained both exactly and by different approximation schemes which are then discussed thoroughly.
		This includes the introduction of an approximation of the polylogarithm difference which is capable of accessing a parameter range neither covered by Sommerfeld expansion nor by Boltzmann approximation.
		It is found that the Fermi temperature and several other temperature scales may be very low, giving rise to experimentally observable behaviours differing from the one described by Fermi liquid theory.
		In particular, two kinds of apparent Fermi liquid behaviour emerge at intermediate temperatures.
		This behaviour is related to recent transport data reported for CuCr${}_{1-\mathrm{x}}$Mg${}_{\mathrm{x}}$O${}_{\mathrm{2}}$ [A. Maignan \textit{et. al}, Solid State Commun.~\textbf{149}, 962 (2009)] and CuRh${}_{1-\mathrm{x}}$Mg${}_{\mathrm{x}}$O${}_{\mathrm{2}}$ [A. Maignan \textit{et. al}, Phys.~Rev.~B~\textbf{80}, 115103 (2009)] by means of the temperature independent correlation functions ratio approximation.
		In this way an effective density of states as well as the effective charge carrier density of these materials are determined.
		Furthermore, conclusions about the specific heat of the latter material are drawn which presents particular effects of the analytical anomaly.
	}}

	\maketitle\thispagestyle{empty}
	\newpage
	\pagestyle{headings}

\section{Introduction}\label{sec:intro}

	Thermoelectric materials provide new opportunities for energy savings with possible usage in a vast area of applications.
	From the fundamental point of view their optimization remains challenging, but several classes of promising materials have been identified.
	They can be manganites \cite{Othaki95}, titanates \cite{Ohta2007}, clathrates \cite{Nolas01,Saramat06} or skutterudites \cite{Tang01}, to quote a few, but also correlated metals, such as Na${}_{\mathrm{x}}$CoO${}_{\mathrm{2}}$ \cite{Terasaki97} or band insulators, such as In${}_{\mathrm{2-x}}$Ge${}_{\mathrm{x}}$O${}_{\mathrm{2}}$ \cite{Berardand08}.
	In the former case their large thermopower was attributed to narrow bands, while sharp band edges seem to play an important role in the latter case, thereby motivating focusing on lower dimensional materials.
	In this context the delafossites \cite{Shannon71,Isawa02}, such as Mg-doped CuCrO${}_{\mathrm{2}}$ \cite{Doumerc86,Maignan091} or Mg-doped CuRhO${}_{\mathrm{2}}$ \cite{Maignan092}, show great promise.
	Even though they were not yet thoroughly studied, their dimensionless figure of merit $ZT$ exceeds 0.1 above $1000\,\mathrm{K}$ \cite{Kuriyama06,Ono07,Hayashi07,Nozaki07}.

\begin{figure}[t!]
		\centering
		\includegraphics[width=\columnwidth]{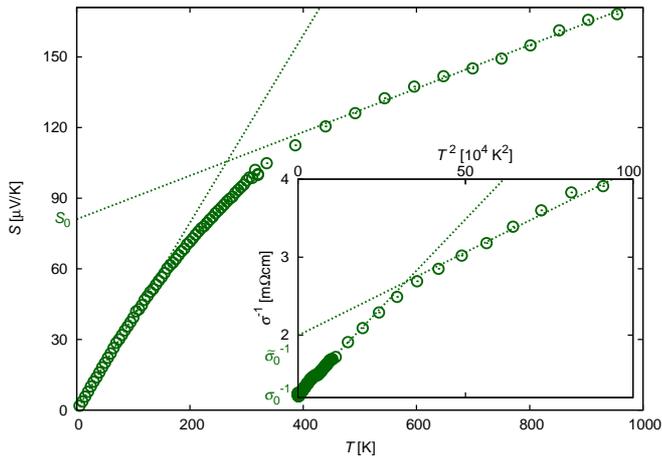}
		\caption{
			Thermopower $S$ and resistivity $\sigma^{-1}$ (inset) of CuRh${}_{0.9}$Mg${}_{0.1}$O${}_2$ as functions of the temperature $T$ (circles).
			The linear, respectively quadratic, regions are highlighted by the dotted lines in order to show the transition from a behaviour, which can be explained within a Fermi liquid picture, towards a behaviour which is referred to as the one of an apparent Fermi liquid (AFL) and characterised by the additional offsets $S_0, {\tilde \sigma}^{-1}_0$.
		}
		\label{fig:AFL}
	\end{figure}

	At those large temperatures usually a temperature independent Heikes behaviour \cite{Heikes61,Chaikin76,Maignan04} is expected.
	Nevertheless, many materials were synthesized showing a $T$-linear thermopower \cite{Okabe03,Singh07,Usui09,Silk09}.
	This behaviour can be explained by Fermi liquid theory, but requires a large temperature scale.
	However, recent findings in manganites \cite{Tanaka96,Fresard02} and delafossites \cite{Maignan092,Miclau07,Maignan091} show slight deviations from Fermi liquid behaviour which challenge this interpretation.
	These modifications include the observation of non-vanishing zero temperature values of the linearly (quadratically) extrapolated thermopower (conductance) and a change in the slope between the low and high temperature behaviour which for instance can be seen in Fig.~\ref{fig:AFL}.
	A structural phase transition might explain such a crossover, but has never been evidenced experimentally.
	In order to study this phenomenon in more detail, it will be referred to an apparent Fermi liquid (AFL) in the following to account for the differences to an ordinary one.

	Although much effort has been devoted to the determination of the electronic structure of the materials where the AFL behaviour was observed \cite{Maignan092,Okabe03,Singh07,Usui09,Maignan091,Felser98,Eyert081,Eyert082} little is known of a microscopic mechanism creating such a behaviour.
	However, a recent study \cite{Maignan092} revealed a dispersionless region in the vicinity of the Fermi energy along a particular direction of the Brillouin zone.
	Conclusively, this leads to a discontinuity in the density of states.
	While on the one hand such features might be difficult to resolve numerically, physical properties on the other are known to be strongly dependent on such analytical anomalies.
	As found for other anomalies, like a cusp in the density of states \cite{Schmitt10} or van-Hove singularities \cite{Varma89,Newns91}, this could render other phases observable, like marginal \cite{Varma89} or non-Fermi liquid \cite{Newns91} ones.
	Therefore in the presence of a discontinuous density of states novel phases should emerge which might explain an AFL behaviour.

	With the purpose of investigating the implications of such a discontinuous density of states, in particular an AFL behaviour, a phenomenological framework is addressed.
	In this framework the chemical potential can be evaluated by the exact solution of the grand canonical system using polylogarithm functions \cite{Lewin81} and applying different approximation schemes to determine the temperature dependence of the chemical potential.
	In order to access the region between Sommerfeld expansion and Boltzmann approximation this includes the introduction of an additional approximation.
	While this approximation starts from the difference  of the polylogarithm functions, it can be interpreted as a way to obtain an averaged density of states and a renormalised doping for a given temperature even for general density of states.
	Based on the chemical potential of the model density of states, the mechanism allows to identify the various temperature ranges associated to the different behaviours.
	This includes not only the Fermi liquid behaviour, but also two types of AFL behaviour with different apparent effective masses.
	The evaluation of the thermopower will be performed using the temperature independent correlation functions ratio approximation (TICR) \cite{Maignan092,Fresard02}.
	It enables to access this quantity within the phenomenological approach in contrast to the ones starting from model Hamiltonians, like the high frequency formulation of transport coefficients \cite{Shastry06p,Peterson07} or the dynamical mean field theory \cite{Pruschke93p,Palsson98}.
	Additionally, the method, generalizing modified Heikes formulas \cite{Chaikin76,Marsh96,Koshibae00,Koshibae01}, can account for some disorder and interaction effects, contrary to approaches using the constant scattering time approximation within Boltzmann transport theory \cite{Maignan092,Singh07,Usui09,Silk09,Wilson-Short07}.
	Furthermore, the TICR approximation allows to discuss the thermopower when governed by its thermodynamic part, relating it to the chemical potential.
	In order to gain further knowledge of the new phases, the model is then applied to CuCr${}_{1-\mathrm{x}}$Mg${}_{\mathrm{x}}$O${}_{\mathrm{2}}$ \cite{Maignan091} and CuRh${}_{1-\mathrm{x}}$Mg${}_{\mathrm{x}}$O${}_{\mathrm{2}}$ \cite{Maignan092}.
	Thereby extracting not only the parameters of the density of states out of the thermopower, but accessing the effective charge carrier density as well.
	The obtained results are then compared to the ones given by the calculation using first principle studies and by statistical electron diffraction spectroscopy (EDS).

	The structure of this investigation concerning the relation of an AFL behaviour and analytic anomalies starts with the outline of the general framework in Sect.~\ref{sec:overview}.
	Therein the form of the used effective density of states will be motivated for the delafossites in more detail and effective variables are defined.
	In the next part, Sect.~\ref{sec:theory}, theoretical background calculations are given in order to determine the phase diagram based on the behaviours found for the chemical potential.
	Therefore it contains the discussion of different approximation schemes.
	On the one hand Sect.~\ref{sec:sommerfeld} places Sommerfeld expansion in the framework of polylogarithm and exemplifies the determination of characteristic temperature scales, while on the other hand the results of Boltzmann approximation are pointed out in Sect.~\ref{sec:boltzmann} and are prepared for the observation of an AFL behaviour within this regime.
	In addition a third approximation examining the region between these two by the means of the approximation of the polylogarithmic difference (APLD) is introduced therein, too (Sect.~\ref{sec:apld}).
	It is then followed by Sect.~\ref{sec:application} including a detailed discussion of the TICR framework and the implication to the thermopower of these approximation schemes (Sect.~\ref{sec:ticr}) as well as the application of the approach to the thermopower data of CuCr${}_{1-\mathrm{x}}$Mg${}_{\mathrm{x}}$O${}_{\mathrm{2}}$ (Sect.~\ref{sec:cucro2}) and CuRh${}_{1-\mathrm{x}}$Mg${}_{\mathrm{x}}$O${}_{\mathrm{2}}$ (Sect.~\ref{sec:curho2}).
	The specific heat following from the obtained effective density of states of the latter material is discussed therein, too.
	The results are summarized in Sect.~\ref{sec:conclusion}.

\section{General Overview}\label{sec:overview}

	In order to find a suitable approach for the delafossites showing an AFL behaviour their band structure should be considered.
	First principle studies \cite{Maignan091,Singh07,Eyert081} showed a sharp increase in the density of states at the Fermi energy close to the upper band edge.
	Since these investigations did not focus on the analytical behaviour of this quantity, possible discontinuities and their implications were not addressed.
	A more recent examination of CuRh${}_{1-\mathrm{x}}$Mg${}_{\mathrm{x}}$O${}_{\mathrm{2}}$ indeed pointed out non dispersive bands in particular directions in the vicinity of the Fermi energy \cite{Maignan092}.
	Therein it was proposed that this could be the origin of degenerate states causing the AFL behaviour.
	This motivates investigating a density of states which is dominated by a discontinuity $\rho_0$ at the band edge.
	Additional terms as appearing in a Taylor series\footnote{In fact, the calculation can easily be generalized to arbitrary (non-integer) powers in the series by replacing the factorial by the Gamma function. From this procedure complications only arise when taking into account a finite band width $W$ or in the low temperature expansion.} of a single band
	\begin{equation}
		\rho(\varepsilon) = \sum_{n=0}^{\infty} \frac{\rho_n}{n!} \varepsilon^n \cdot \Theta(\varepsilon) \Theta(W-\varepsilon)\,,\label{eq:Taylorrho}
	\end{equation}
	where $W$ denotes the band width, should therefore be small corrections.
	With such a density of state an AFL behaviour should accurately be described since at the observed large temperatures the chemical potential is believed to be located in the band gap.
	Then the product of the Fermi function with the density of states changes mostly at the band edge.
	In the finite temperature calculations following, mainly a sufficiently large band is assumed, i.e. \mbox{$W \rightarrow \infty$}, while effects of a finite band width will only be noted briefly.
	In order to exemplify perturbations to the case where Eq.~(\ref{eq:Taylorrho}) only includes a discontinuity, results will often be visualized in the case where only the first two terms in the sum of Eq.~(\ref{eq:Taylorrho}) are present, i.e. \mbox{$\rho(\varepsilon) = (\rho_0 + \rho_1\varepsilon)\cdot\Theta(\varepsilon)$}.

	From this density of states the doping $\mathrm{x}$ can be analytically calculated for a given chemical potential $\mu$ as
	\begin{equation}
		\mathrm{x} = \int_{-\infty}^{\infty}\mathrm{d}\varepsilon \, \big(1 - f(\varepsilon -\mu)\big)\rho(\varepsilon)\,,\label{eq:doping}
	\end{equation}
	where $f(x)$ represents the Fermi function.
	The integration occurring in Eq.~(\ref{eq:doping}) in combination with Eq.~(\ref{eq:Taylorrho}) can be performed using polylogarithm functions \cite{Lewin81}:
	\begin{align}
		y = &-\sum_n r_n \beta^{-(n+1)} \Bigg[\mathrm{Li}_{n+1}\big(\nu(\beta)\big) \nonumber\\
		&-  \sum_{j=0}^n (\beta W)^{n-j} \frac{1}{(n-j)!} \,\mathrm{Li}_{j+1}\big(\nu(\beta)\mathrm{e}^{-\beta W}\big) \Bigg]\,,\label{eq:exact}
	\end{align}
	where the polylogarithm functions
	\begin{align}
		\mathrm{Li}_{n+1}(x^{-1})=\frac{1}{n!}\int_0^{\infty}\frac{t^{n}}{x\exp(t)-1}\,\mathrm{d} t\nonumber
	\end{align}
	as well as the renormalised values and the negative fugacity,
	\begin{equation}
		y=\frac{\mathrm{x}}{\rho_0}, \qquad r_n=\frac{\rho_n}{\rho_0}, \qquad \nu(\beta)=-\mathrm{e}^{\beta\mu}\,,\label{eq:yrnu}
	\end{equation}
	were introduced.
	The exact solution of Eq.~(\ref{eq:exact}) would give the temperature dependence of the chemical potential.
	Unfortunately, it can only be performed numerically, but useful insight can be gained by different approximation schemes.
	Those will be discussed in the following section.

\section{Theoretical Background}\label{sec:theory}

  \subsection{Sommerfeld Region}\label{sec:sommerfeld}

	From Equation (\ref{eq:exact}) the result of Sommerfeld expansion can easily be reproduced:
	By expanding the polylogarithm functions for large negative arguments \cite{Lewin81}, respectively low temperatures (\mbox{$\beta\mu \gg 1$}), up to the first polynomial terms in temperature,
	\begin{align}
		\mathrm{Li}_n(-\mathrm{e}^{-\beta\mu})
		&= 2\sum_{m=0}^{\lfloor n/2 \rfloor} \mathrm{Li}_{2m}(-1)\, \frac{(-\beta\mu)^{n-2m}}{(n-2m)!}\nonumber\\
		&~~~- (-1)^n\, \mathrm{Li}_n\!\left(-\mathrm{e}^{\beta\mu}\right)\\
		&= -\frac{1}{n!}\,(-\beta\mu)^{n} - \frac{\pi^2}{6} \frac{n(n-1)}{n!}\,(-\beta\mu)^{n-2}\nonumber\\
		&~~~+ \mathcal{O}\big((\beta\mu)^{n-4}\big) + \mathcal{O}\big(\mathrm{e}^{\beta\mu}\big)\,,\label{eq:Liexpansion}
	\end{align}
	a quadratic equation for the chemical potential is obtained, if the terms of order $\mathcal{O}\big(\beta^{-4}\big)$, $\mathcal{O}\big(\mathrm{e}^{\beta\mu}\big)$ and the effects of a finite band width $W$ are neglected.
	Expanding its solution for small temperatures yields the familiar result of Sommerfeld expansion:
	\begin{equation}
		\mu_S = k_{\mathrm{B}}T_F \left( 1 - k_{\mathrm{B}}{}^{-1}T_F\varpi \cdot \left( \frac{T}{T_F} \right)^2\right)\,,\label{eq:sommerfeld}
	\end{equation}
	where the Fermi liquid parameter $\varpi$ and the Fermi temperature $T_F$ are given by
	\begin{equation}
		\varpi = \frac{\pi^2}{6}k_{\mathrm{B}}{}^2\frac{\rho'(T_F/k_{\mathrm{B}})}{\rho(T_F/k_{\mathrm{B}})} \quad \mathrm{and} \quad T_F \approx \frac{\sqrt{1 + 2yr_1} - 1}{k_{\mathrm{B}} r_1}\,, \label{eq:fermitemp}
	\end{equation}
	with $\rho'(\varepsilon)$ denoting the first derivative of the density of states.
	For the latter result was assumed that the coefficients in the Taylor series, Eq.~(\ref{eq:Taylorrho}), are decreasing with growing order, i.e. $r_n \ll r_1$ for $n > 1$.

\begin{figure*}[t!]
	\centering
	\includegraphics[width=\textwidth]{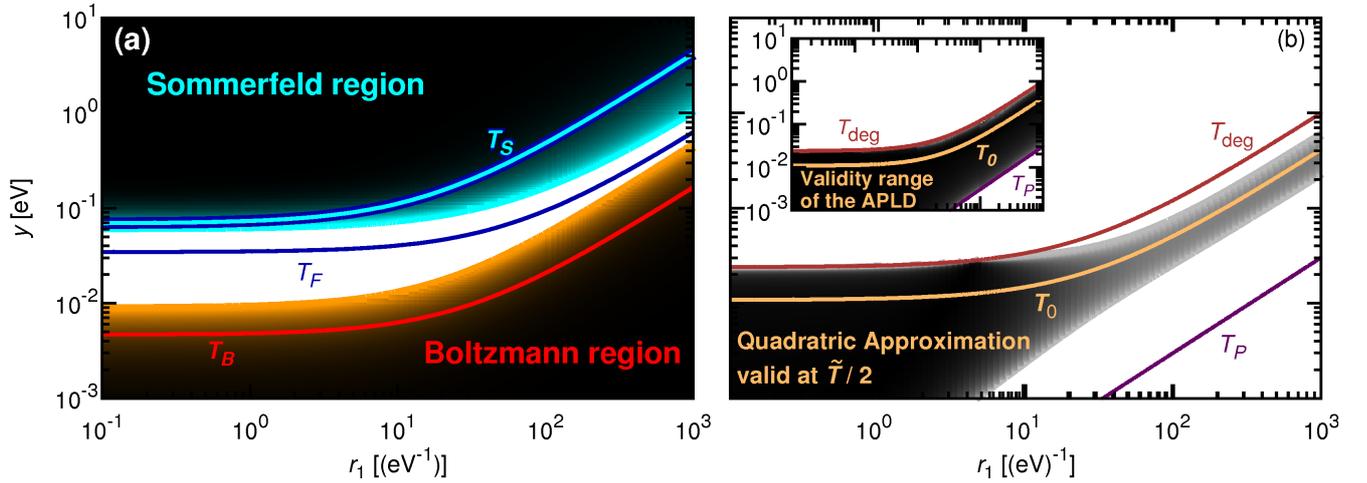}
	\caption{
		Relative errors of the chemical potential obtained by the discussed approximations to the one calculated numerically from Eq.~(\ref{eq:exact}) as functions of the parameter ratio $r_1$ and the renormalised doping $y$ for a specific temperature of \mbox{$T=400\,\mathrm{K}$} and a density of states given by $\rho(\varepsilon)=(\rho_0+\rho_1\varepsilon)\,\Theta(\varepsilon)$.
		The shading in Fig.~\ref{fig:valrange}a runs from 0\% (darkest) to 10\% (lightest blue) for the comparison of the solution from Sommerfeld expansion Eq.~(\ref{eq:sommerfeld}), and from 0\% (darkest) to 10\% (lightest orange) for the Boltzmann approximation Eq.~(\ref{eq:boltzmannnoapprox}), respectively 0\% to 100\% for the one using the APLD result Eq.~(\ref{eq:apldnoapprox}) in the inset of Fig.~\ref{fig:valrange}b.
		The main figure~\ref{fig:valrange}b shows the relative errors for the quadratic expansion of the APLD Eqs.~(\ref{eq:apldrenormalizationstart}) and (\ref{eq:apldrenormalizationend}) with the used expansion temperature \mbox{${\tilde T}=800\,\mathrm{K}$}.
		Additionally drawn are the curves where the specified temperature $T$ matches the Sommerfeld breakdown temperature $T_S$, the Fermi temperature $T_F$, the Boltzmann breakdown temperature $T_B$, the degeneracy temperature $T_{\mathrm{deg}}$, the temperature $T_P$, where the APLD has a logarithmic singularity, or the transition temperature $T_0$, where the chemical potential equates with the negative thermal energy:\mbox{$\mu = -k_{\mathrm{B}}T_0$}.
		In the inset of (b) the degeneracy temperature $T_{\mathrm{deg}}$ as well as $T_P$ are very close to the points where the relative error between the APLD and the numerical solution increases rapidly.
	}
	\label{fig:valrange}
\end{figure*}
	When Sommerfeld expansion is applied to determine the low temperature behaviour, the convergence of the Taylor series used therein is usually not considered.
	For a Fermi energy close to a discontinuity this induces a further breakdown criterion.
	It can be obtained when recalling that the expansion~(\ref{eq:Liexpansion}) is made in the case $\beta\mu \gg 1$.
	As a measure of the condition where this expansion is violated, the temperature where $\beta\mu = 2$ can therefore be taken.
	With this the criterion the expansion~(\ref{eq:Liexpansion}) shows that the Fermi liquid expression should be valid up to the temperature
	\begin{equation}
		T_S = \frac{\sqrt{1 + T_F\varpi/k_{\mathrm{B}}} - 1}{\varpi/k_{\mathrm{B}}} \,.\label{eq:sommerfeldtemp}
	\end{equation}
	Therefore the neglected terms in this approximation applied to Eq.~(\ref{eq:yrnu}) are smaller than \mbox{$\nu^{-1}=-\mathrm{e}^{-2}$}.
	While the temperature $T_S$ will always be lower than the Fermi temperature $T_F$, the comparison to the numerical solution of Eq.~(\ref{eq:exact}) in Fig.~\ref{fig:valrange}a shows that it seems to be important only when the effect of the discontinuity in the density of states is more important than that of the slope.
	Accordingly, in the case of a dominating discontinuity $T_S$ describes the numerically studied breakdown of the Fermi liquid more accurately.
	In addition, the breakdown temperature $T_S$ and the Fermi temperature $T_F$ as well as the numerical solution show for large renormalised doping $y$ and small parameter ratios $r_n$ a region where Sommerfeld expansion is valid at room temperature, and even beyond.

  \subsection{Boltzmann Region}\label{sec:boltzmann}

	On the opposite side of physical temperatures, the result of Boltzmann approximation can be deduced as a Taylor expansion of the polylogarithm functions in $\nu$ in the limit $\beta\mu \ll -1$.
	The result in first order is given by
	\begin{equation}
		\mu_B = -k_{\mathrm{B}}T\,\ln\!\left(\frac{k_{\mathrm{B}}T\, {\tilde r}(k_{\mathrm{B}}T)}{y}\right)\,,\label{eq:boltzmannnoapprox}
	\end{equation}
	where the quantity ${\tilde r}(\varepsilon) = \sum_n r_n\varepsilon^n$ was introduced which will be discussed in the following section \ref{sec:apld}.
	In case the band width $W$ is taken into account the defined quantity is shifted in this approximation to \mbox{${\tilde r}(\varepsilon) - \sum_n r_n \varepsilon^n \Gamma(n+1,W/\varepsilon)/n!$}, where $\Gamma(n,x)$ denotes the upper incomplete gamma function.
	Since this function becomes an exponentially decreasing function for large second argument, i.e. $\Gamma(n,x) \rightarrow \mathrm{e}^{-x}$ for $x \rightarrow \infty$, the effects of a finite band width can be neglected if the thermal energy $k_{\mathrm{B}}T$ is still smaller than the band width.

	In case the band width is neglected a similar break down temperature as in Eq.~(\ref{eq:sommerfeldtemp}) can be extracted for Eq.~(\ref{eq:boltzmannnoapprox}) but for the criterion $\beta\mu = -2$.
	This means the expression~(\ref{eq:boltzmannnoapprox}) should be valid for temperatures larger than
	\begin{equation}
		T_B \approx \frac{\sqrt{1 + 4yr_1\mathrm{e}^{2}} - 1}{2k_{\mathrm{B}}r_1}\,,\label{eq:boltzmanntemp}
	\end{equation}
	where corrections from this approximation applied to Eq.~(\ref{eq:exact}) are of order \mbox{$\nu^2=\mathrm{e}^{-4}$} and where again a decreasing importance of the Taylor coefficients with growing order was assumed.

	In order to bring the expression (\ref{eq:boltzmannnoapprox}) into a similar form as obtained in the low temperature expansion it can be linearised
	around a certain temperature ${\tilde T}$.
	However, it will then show a finite linear term in contrast to Eq.~(\ref{eq:sommerfeld}):
	\begin{align}
		\mu_B &= \mu_B^{(0)} + \mu_B^{(1)} \cdot (T-{\tilde T}) + \frac{1}{2} \mu_B^{(2)} \cdot (T - {\tilde T})^2 \quad \mathrm{with}\label{eq:boltzmann}\\
		\mu_B^{(0)} &= -k_{\mathrm{B}} {\tilde T}\, \ln \frac{k_{\mathrm{B}}{\tilde T}\, {\tilde r}(k_{\mathrm{B}}{\tilde T})}{y}\,,\\
		\mu_B^{(1)} &= -k_{\mathrm{B}} \ln \frac{ k_{\mathrm{B}}{\tilde T}\, {\tilde r}(k_{\mathrm{B}} {\tilde T})}{y} - k_{\mathrm{B}} [1+k_{\mathrm{B}}{\tilde T}\, (\ln{\tilde r})^{(1)}]\,,\label{eq:boltzmannlin}\\
		\mu_B^{(2)} &= -\frac{2k_{\mathrm{B}}}{\tilde T}[1 + 2k_{\mathrm{B}} {\tilde T}\, (\ln{\tilde r})^{(1)} + (k_{\mathrm{B}}{\tilde T})^2\,(\ln{\tilde r})^{(2)}]\,,\label{eq:boltzmannquad}
	\end{align}
	where $(\ln{\tilde r})^{(n)}$ denotes the $n$-th derivative of the logarithm of the quantity ${\tilde r}$ at the expansion energy $k_{\mathrm{B}}{\tilde T}$.
	These terms are not described within the Fermi liquid picture, establishing instead a region of an apparent Fermi liquid (AFL).
	In particular, for an AFL behaviour a non-vanishing linear term in the chemical potential is obtained which even does not vanish when reordering the coefficients like in the Fermi liquid case
	\begin{align}
		\mu_B = &\big(\mu_B^{(0)} - \mu_B^{(1)}{\tilde T} + \mu_B^{(2)}{\tilde T}^2/2\big)\nonumber\\
		&+ \big(\mu_B^{(1)} - 2\mu_B^{(2)}{\tilde T}\big) \cdot T + \mu_B^{(2)}/2 \cdot T^2\,.\nonumber
	\end{align}
	However, due to the expansion around a certain temperature the coefficients depend on this quantity, too.
	Nevertheless, for large temperatures the terms in squared brackets in Eqs.~\ref{eq:boltzmannlin} and~\ref{eq:boltzmannquad} are constant in leading order, resembling only a weak dependence of the coefficients.

  \subsection{Intermediate Region}\label{sec:apld}

	In order to fill the gap between Boltzmann and Sommerfeld region, it is useful to take a closer look at the similarity of the polylogarithm of different orders.
	To that aim it is useful to recall that every polylogarithm behaves similar at small negative arguments, e.g. $\mathrm{Li}(0)=0$, $\mathrm{Li'}(0)=1$.
	Since the first polylogarithm resembles an elementary function \mbox{$\mathrm{Li}_1(\nu) = -\ln(1-\nu)$}, it is very appealing to consider the other orders in terms of an approximation of the polylogarithm difference (APLD)
	\begin{equation}
		d_n\!(\nu) = \mathrm{Li}_{n+1}(\nu) - \mathrm{Li}_1(\nu)\,,
	\end{equation}
	occurring in Eq.~(\ref{eq:exact}), in terms of a Taylor series around a negative expansion fugacity $\nu_0$:
	\begin{align}
		y &= \beta^{-1} {\tilde r}\big(\beta^{-1}\big)\ln(1-\nu) -\sum_{n\not = 0} r_n d_n\!(\nu_0) \,\beta^{-(n+1)}\nonumber\\
		   &~~~-\sum_{n\not = 0} r_n \big(d_{n-1}(\nu_0) - d_{-1}(\nu_0)\big)\,\nu_0{}^{-1}\,\beta^{-(n+1)}\,(\nu - \nu_0)\nonumber\\
		   &~~~- \mathcal{O}\big((\nu-\nu_0)^2\big)\,, \label{eq:diff}
	\end{align}
	where for simplicity again the limit $W\rightarrow \infty$ was considered.

	The first terms, proportional to the logarithm, in this approximation can be understood as following from a temperature dependent, averaged density of states.
	Reviewing the expressions of the previous subsection it has already occurred as the denominator of the fugacity in the result of Boltzmann approximation in Eq.~(\ref{eq:boltzmannnoapprox}).
	In the case of a dominant discontinuity or slope it takes the form
	\begin{equation}\label{eq:rrenormalized}
		{\tilde r}(\varepsilon) \approx \rho(k_{\mathrm{B}}T) / \rho_0\,,
	\end{equation}
	but provides different expressions for more general density of states.
	Additionally, the zeroth order of the Taylor series only shifts the doping value used in the theory
	\begin{equation}\label{eq:dx}
		y \rightarrow y + dy(k_{\mathrm{B}} T) \quad\mathrm{with}\quad dy(\varepsilon)=\sum_{n\not = 0} r_n\, d_n\!(\nu_0) \,\varepsilon^{n+1}\,,
	\end{equation}
	since this term does not depend on the chemical potential.
	In contrast, the first order terms in this expansion would result in the chemical potential given by
	\begin{align}\label{eq:apld1st}
		&k_{\mathrm{B}} T\,\bar{r}\,(1 + \mathrm{e}^{\beta\mu})\exp\!\left( k_{\mathrm{B}} T\,\bar{r}\,(1 + \mathrm{e}^{\beta\mu}) \right) \nonumber\\
		&~= k_{\mathrm{B}} T\,\bar{r}\exp\!\left( \frac{y+dy(k_{\mathrm{B}} T) - k_{\mathrm{B}} T\,\bar{r}\,(\nu_0 - 1)}{k_{\mathrm{B}} T\,{\tilde r}(k_{\mathrm{B}} T)}\right)\!,
	\end{align}
	where $k_{\mathrm{B}} T\bar{r}$ is the prefactor of the linear term in Eq.~(\ref{eq:diff}) similar to $dy(k_{\mathrm{B}} T)$ for the zeroth order.
	The solution of this equation can be found as the Lambert~$\mathcal{W}$ function \cite{Corless96} with the right-hand side as argument.
	Of course, for small (positive) arguments this function can again be linearised.
	Since this leads to the same solution as if only the zeroth order term is taken, the first order contributions only give corrections at large doping, Eq.~(\ref{eq:dx}), or low temperatures.
	Therefore this as well as all higher orders will be neglected in the following.
	The chemical potential in this approximation is then given by
	\begin{align}
		\mu_P &= k_{\mathrm{B}} T\,\ln\!\left[\exp\!\left( \frac{y + dy(k_{\mathrm{B}} T)}{k_{\mathrm{B}} T\,{\tilde r}(k_{\mathrm{B}} T)}\right) - 1 \right]\label{eq:apldnoapprox}\\
		      &= k_{\mathrm{B}} T\,\big(z + \ln(2\sinh z)\big)\,,\label{eq:apldnoapprox2}
	\end{align}
	with the argument of the hyperbolic function
	\begin{equation}
		z = \frac{y+dy(k_{\mathrm{B}} T)}{2k_{\mathrm{B}} T\,{\tilde r}(k_{\mathrm{B}} T)} = \frac{y+\sum_{n\not = 0} r_n \,d_n\!(\nu_0) \,(k_{\mathrm{B}} T)^{n+1}}{2\sum_n r_n\, (k_{\mathrm{B}} T)^{n+1}}\,.\label{eq:hyparg}
	\end{equation}
\par

	The validity of this expression can be related to the results of Boltzmann approximation and Sommerfeld expansion in particular cases:
	For a very large expansion temperature the difference of the polylogarithm $d_n(\nu_0)$ becomes negligible.
	Due to the assumed high temperatures the result of Boltzmann approximation $\mu_B$ in Eq.~(\ref{eq:boltzmannnoapprox}) is obtained by an additional expansion of the exponential function in Eq.~(\ref{eq:apldnoapprox}).
	Contrastively, in decreasing the expansion temperature the fugacity will reach unity, i.e. $\nu_0 \rightarrow -1$.
	After this point a slight variation of the chemical potential or the temperature will lead to a greater change in the fugacity.
	Therefore higher orders become important and the Taylor series in Eq.~(\ref{eq:diff}) will break down for smaller temperature differences.
	This means the APLD is only appropriate for temperatures larger than the degeneracy temperature $T_{\mathrm{deg}}$ where the chemical potential vanishes.
	According to Eq.~(\ref{eq:exact}) this temperature is implicitly given by
	\begin{equation}
		y = -\sum_n r_n \,\mathrm{Li}_{n+1}(-1)\,(k_{\mathrm{B}} T_{\mathrm{deg}})^{n+1}\,.\label{eq:degtempgen}
	\end{equation}
	If the Taylor series in Eq.~(\ref{eq:Taylorrho}) is dominated by its first terms, it can be solved yielding
	\begin{equation}
		T_{\mathrm{deg}} \approx \frac{-\ln 2 + \sqrt{(\ln 2)^2 + \pi^2yr_1/3}}{\pi^2k_{\mathrm{B}}r_1/6}\,.\label{eq:degtemp}
	\end{equation}
	However, for a very dominant discontinuity, the first term in Eq.~(\ref{eq:diff}) will give the dominant contribution to the doping.
	Therefore, the other terms will resemble small corrections whose importance will decrease as the first term increases.
	Since the latter was treated exactly, this means that the validity region is increased to very low temperatures.
	In particular, for the case the density of states is only described by a discontinuity the APLD becomes exact and resembles the result of Sommerfeld expansion after performing a Taylor series for small temperatures.

	Surprisingly, at finite values of the ratios $r_n$ and the expansion parameters $d_n(\nu_0)$ the chemical potential in this approximation has a logarithmic singularity where the argument of the exponential function in Eq.~(\ref{eq:apldnoapprox}) vanishes.
	In order to avoid this divergence Boltzmann approximation should be used for temperatures similar or larger than this temperature $T_P$ which is given for the first two terms of the Taylor series in Eq.~(\ref{eq:Taylorrho}) by
	\begin{equation}
		T_P \approx \sqrt{\frac{y}{-d_1(\nu_0) \,r_1k_{\mathrm{B}}{}^2}}\,. \label{eq:tp}
	\end{equation}

	A suitable expansion point which lies between the Sommerfeld and Boltzmann breakdown temperatures given by Eqs.~(\ref{eq:sommerfeldtemp}) and~(\ref{eq:boltzmanntemp}) is the transition temperature $T_0$ where the exact chemical potential, as given by Eq.~(\ref{eq:exact}), is equal to the negative thermal energy \mbox{$\mu = -k_{\mathrm{B}}T$} or \mbox{$\nu_0=-\mathrm{e}^{-1}$}.
	Similar to Eq.~(\ref{eq:degtemp}) this temperature can be stated explicitly if the Taylor series in Eq.~(\ref{eq:Taylorrho}) is dominated by the first terms.
	In this case it is given by
	\begin{equation}
		T_0 \approx \frac{\ln(1-\nu_0) - \sqrt{\big(\ln(1-\nu_0)\big)^2 - 4yr_1\,\mathrm{Li}_{2}(\nu_0)}}{2k_{\mathrm{B}}r_1\,\mathrm{Li}_{2}(\nu_0)}\,. \label{eq:t0}
	\end{equation}
	With this expansion temperature $T_0$ further simplifications can be brought to Eq.~(\ref{eq:apldnoapprox}):
	In this case the absolute value of the difference of the polylogarithm functions of first orders amounts to $|d_1\!(\nu_0)|=0.03$ and monotonically decreases with larger expansion temperatures.
	Multiplying this number by the typical values for $k_{\mathrm{B}}{}^2T^2$ at room temperature leads to a result which is much smaller than the renormalised doping $y$.
	Although the difference increases slightly for higher orders, the shift of the doping $dy$ in the result of the APLD, Eq.~(\ref{eq:apldnoapprox}), may be neglected, especially if the Taylor series is governed by its first terms.
	Since this means neglecting even the first order in the expansion of the APLD (\ref{eq:diff}) the expansion point $\nu_0$ would not be present in the solution of the chemical potential Eq.~(\ref{eq:apldnoapprox}).
	Therefore an even larger application region is expected in this case which is in agreement with the diverging behaviour of the break down temperature $T_P$ as well as with the discussion of the lower break down temperature after Eq.~(\ref{eq:degtemp}).
\par

	Since in this limit the results of the APLD can be understood as those following from a constant averaged density of states given in Eq.~(\ref{eq:rrenormalized}), at first the chemical potential of only a discontinuity will be studied, i.e. \mbox{$\rho(\varepsilon) = \rho_0\,\Theta(\varepsilon)\Theta(W-\varepsilon)$}.
	In this limit, where the APLD becomes exact, a finite bandwidth $W$ shifts the chemical potential by
	\begin{equation}
		\mu_P \rightarrow \mu_P -k_{\mathrm{B}} T\,\ln\!\left[1-\exp\!\left( \frac{y-W}{k_{\mathrm{B}} T\,{\tilde r}(k_{\mathrm{B}} T)}\right)\right]\,,
	\end{equation}
	which resembles a similar contribution as Eq.~\ref{eq:apldnoapprox} but for the upper band edge.
	Following from this shift the chemical potential reaches a constant value in the high temperature limit $\beta \rightarrow 0$,
	\begin{equation}\label{eq:muHeikes}
		\mu^{(-)}_P \rightarrow k_{\mathrm{B}}T\,\ln\left(\frac{y}{W - y}\right)\,,
	\end{equation}
	in contrast to the chemical potential obtained by Boltzmann approximation in Eq.~(\ref{eq:boltzmannnoapprox}).
	Therefore, the assumption of that approximation $\beta\mu \ll -1$ is not valid when the thermal energy surpasses the band width $W$.
	However, for a sufficiently large band width $W \gg k_\mathrm{B}T + y$ with respect to the temperature one is interested in, the shift can still be neglected.

	\begin{figure}[t!]
		\centering
		\includegraphics[width=\columnwidth]{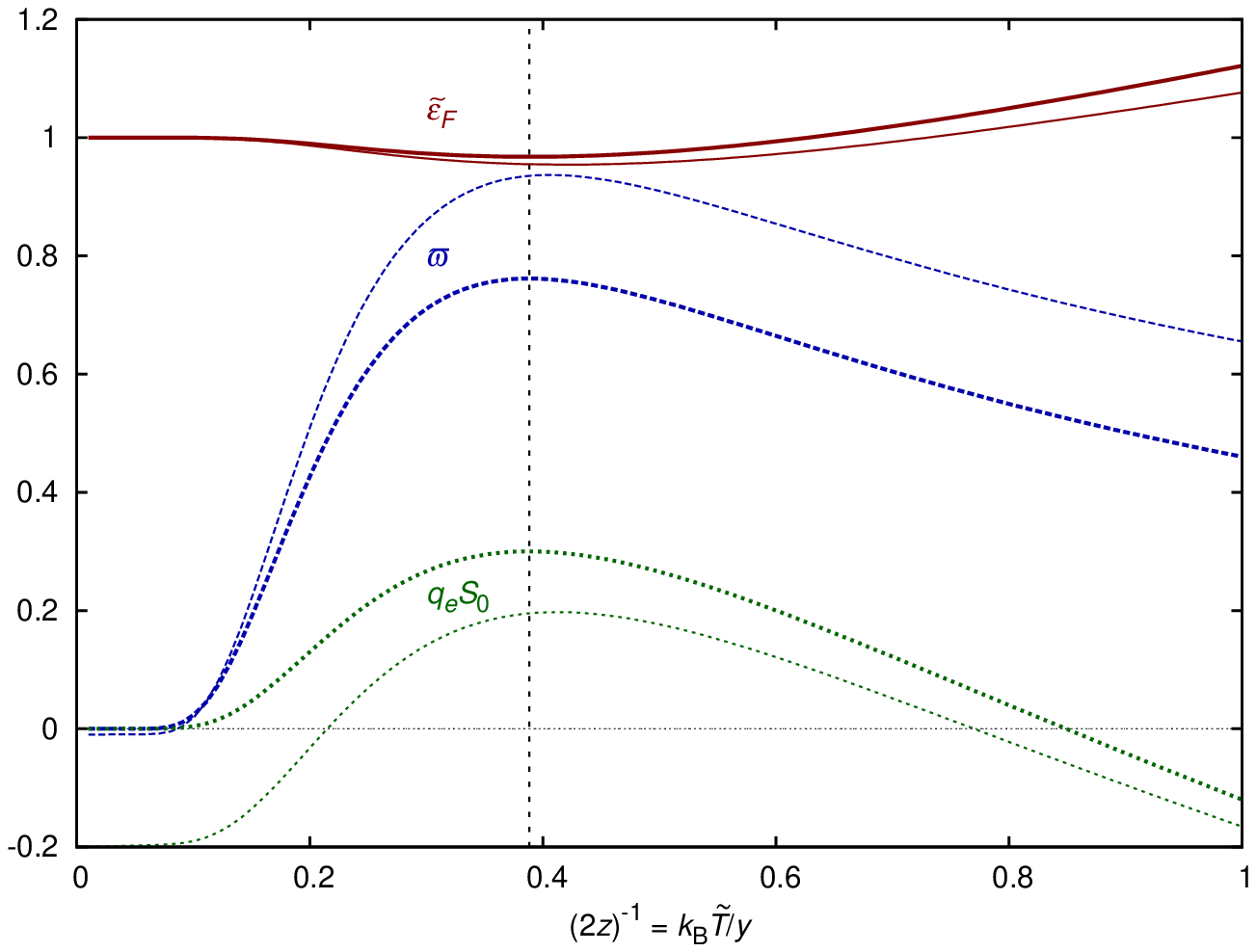}
		\caption{
			Dependence of the coefficients of the quad\-rat\-ic expansion of the chemical potential in APLD \mbox{$\mu^{(-)}_{P} = {\tilde \varepsilon}_F - q_eS_0 T + \varpi T^2$} on the expansion temperature $\tilde T$ for fixed renormalised doping $y=0.2\,\mathrm{eV}$ and parameter ratio $r_1$ of different regimes.
			For small parameter ratio $r_1$ as given by Eq.~(\ref{eq:apldsmallr}) (thick lines), the temperature dependence of these parameters are weakest around the temperature ${\tilde T}_P^{(-)}$ given by Eq.~(\ref{eq:zmaximum}) (dashed vertical line).
			For significant parameter ratio $r_1=2\,\mathrm{(eV)^{-1}}$ ($r_n=0$ for $n>1$) according to Eqs.~(\ref{eq:apldrenormalizationstart}) and (\ref{eq:apldrenormalizationend}) (thin lines), the stationary points have increased and have begun to shift independently, causing the AFL to destabilize.
			In the figure ${\tilde \varepsilon}_F$, $q_eS_0$, $\varpi$ are given in units of $y^{-1}$, $k_{\mathrm{B}}{}^{-1}$ and $y\,k_{\mathrm{B}}{}^{-2}$ respectively.
		}
		\label{fig:coefficientssmallr}
	\end{figure}
	In order to compare the observed behaviour to the one known from a Fermi liquid as derived in Eq.~(\ref{eq:sommerfeld}), the chemical potential of Eq.~(\ref{eq:apldnoapprox2}) can be brought into a similar form.
	Performing a quadratic expansion around the temperature ${\tilde T}$ leads to
	\begin{align}
		\mu^{(-)}_{P} &= \mu^{(0)}_P\!({\tilde T}) + \mu^{(1)}_P\!({\tilde T}) \cdot (T-{\tilde T})
		     +\frac{1}{2}\mu^{(2)}_P\!({\tilde T}) \cdot (T - {\tilde T})^2\nonumber\\
		    &= -k_{\mathrm{B}}\left[z + \ln(2\sinh z)\right]{\tilde T}\nonumber\\
		     &~~~ +k_{\mathrm{B}}\left[ z\coth z - \ln(2\sinh z)\right] \, (T - {\tilde T})\nonumber\\
		     &~~~ +k_{\mathrm{B}}\frac{z^2}{2\sinh^2 z}\frac{1}{{\tilde T}} \,(T - {\tilde T})^2\,,\label{eq:apldsmallr}
	\end{align}
	with \mbox{$z=y\,(2k_{\mathrm{B}}{\tilde T})^{-1}$}, the argument of the hyperbolic functions as in Eq.~(\ref{eq:hyparg}).
	Due to the fact that the linear coefficient after reordering will not vanish, which can be seen in Fig.~\ref{fig:coefficientssmallr}, this leads to an AFL behaviour.
	Furthermore, since the quadratic coefficient \mbox{$\frac{1}{2}\mu^{(2)}_P\!({\tilde T})$} possesses in its $\tilde T$ dependence an inflexion point near its maximum at
	\begin{equation}
		z\coth z = \frac{3}{2}\label{eq:zmaximum}\,,
	\end{equation}
	the terms of higher order will have small values near this maximum.
	Therefore the quadratic expansion will approximate the function very well around this point defining a particular expansion temperature.
	Since the numerical solution of Eq.~(\ref{eq:zmaximum}) reads \mbox{$z \approx 1.29$} this temperature is given by
	\begin{equation}
		{\tilde T}_P^{(-)} \approx 4505\,\frac{\mathrm{K}}{\mathrm{eV}}\cdot y = 0.8 \,T_S = 0.3\,T_{\mathrm{deg}}\,.\label{eq:expTPlowr}
	\end{equation}

	An interesting aspect of this temperature comes from the fact that the chemical potential have not yet moved far from the Fermi energy.
	In fact, the low temperature result obtained in Eq.~(\ref{eq:sommerfeld}) would describe a chemical potential independent of temperature.
	It therefore remains at the Fermi energy $\varepsilon_F$ with increasing temperature.
	At the specified temperature, when it differs from the Fermi energy by $\big(\varepsilon_F-\mu_P\big({\tilde T}_{P}^{(-)}\big)\big)/\varepsilon_F = 3\%$ it starts moving towards the band edge.
	This is due to the fact that in the doping constraint, Eq.~(\ref{eq:doping}), the Fermi distribution is cut at the band edge where it is reduced by $8\%$ at ${\tilde T}_{P}^{(-)}$.
	This start of the motion of the chemical potential can be approximated as the Taylor series in Eq.~(\ref{eq:apldsmallr}) with non-vanishing quadratic coefficient, therefore distinguishing it from the Fermi liquid parameter.
	Nevertheless, a cross-over from a true Fermi liquid behaviour, i.e.~with finite parameters, to such an AFL can not be observed in this limit.

\begin{figure}[t!]
		\centering
		\includegraphics[width=\columnwidth]{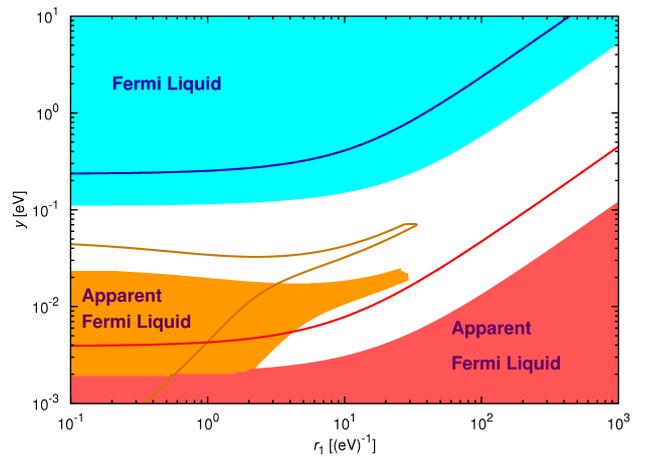}
		\caption{
			Phase diagram for \mbox{$T = 400\,\mathrm{K}$} (areas) and \mbox{$T = 800\,\mathrm{K}$} (lines).
			In the different coloured regions the relative errors between the exact solution of the chemical potential and the one described by the stated behaviour is smaller than 50\%.
			For the AFL by APLD (AFL region on the left), Equations (\ref{eq:apldrenormalizationstart}) and (\ref{eq:apldrenormalizationend}) are compared to Eq.~(\ref{eq:apldnoapprox}), respectively for the one given by Boltzmann approximation (AFL region in the lower right corner) Eq.~(\ref{eq:boltzmann}) to Eq.~(\ref{eq:boltzmannnoapprox}).
			In these regions the more general Fermi liquid and AFL expressions (\ref{eq:sommerfeld}),~(\ref{eq:boltzmannnoapprox}),~(\ref{eq:apldnoapprox}) differ from the exact solution by less than 1\%.
		}
		\label{fig:phasediagram}
	\end{figure}

	However, on the one hand in considering very small additional coefficients of higher order terms in the Taylor series, the real Fermi liquid parameter would be non-vanishing.
	On the other hand, the boost of the chemical potential near the temperature ${\tilde T}_{P}^{(-)}$ may still be
	possible to observe on top of the background behaviour caused by these coefficients.
	Since the breakdown temperature of the observable Fermi liquid $T_S$ will be slightly lower than the temperature where the AFL behaviour occurs, this leads to a comparably small transition region between a Fermi liquid and an AFL behaviour.
	Analytically, this represents itself as follows:
	When taking the effects in Eq.~(\ref{eq:apldnoapprox2}) of the full Taylor-series of the density of states into account, both the hyperbolic argument as described in Eq.~(\ref{eq:hyparg}) and the modification of the derivatives in the quadratic expansion have to be taken into account:
	\begin{align}
		\mu^{(1)}_P\!({\tilde T}) &\rightarrow \mu^{(1)}_P\!({\tilde T}) - k_{\mathrm{B}}{\tilde z}\, (1 + \coth z)\,,\label{eq:apldrenormalizationstart}\\[0.5em]
		\varpi &\rightarrow \varpi \left( 1 + \frac{{\tilde z}}{z}\right)^{\!\!2} +  k_{\mathrm{B}}\frac{1+\coth z}{2\big({\tilde r}(k_{\mathrm{B}}{\tilde T})\big)^{2}}\nonumber\\[0.4em]
		&~~~\cdot\big[z\, k_{\mathrm{B}}{\tilde T}\,{\tilde r}^{(2)} \,{\tilde r}(k_{\mathrm{B}}{\tilde T})  +  dy^{(1)} {\tilde r}^{(1)}\nonumber\\
		&~~~-  2\,z\,k_{\mathrm{B}}{\tilde T}\,\big({\tilde r}^{(1)}\big)^2  -  \frac{1}{2}dy^{(2)}\,{\tilde r}(k_{\mathrm{B}}{\tilde T})\big]\,,\label{eq:apldrenormalizationend}
	\end{align}
	with ${\tilde z} = \big(2\,z\,k_{\mathrm{B}} {\tilde T}\,{\tilde r}^{(1)} - dy^{(1)}\big)/(2{\tilde r})$ and where ${\tilde r}^{(n)}$ denotes the $n$-th derivative of the renormalised density of states and $dy^{(n)}$ respectively the one of the doping change both taken at the thermal energy $k_{\mathrm{B}} {\tilde T}$.
	These adjustments cause the common stationary point of all three coefficients of the chemical potential $\mu_{P}^{(q)}$, as a polynomial in $T$, to shift independently.
	As can be seen in Fig.~\ref{fig:valrange}b, this destabilizes the quadratic approximation thereby destroying the AFL, stressing the role of a large discontinuity for this kind of AFL.
	However, in adjusting the renormalised doping $y$ to given parameter ratios $r_n$, which should not be too large, a narrow region remains where the different stationary points are close to one another.
	As indicated in Fig.~\ref{fig:coefficientssmallr} the observation temperature ${\tilde T}_P^{(q)}$ of the AFL of this region would then be larger than the temperature given in Eq.~(\ref{eq:expTPlowr}).
	Since the degeneracy temperature given by Eq.~(\ref{eq:degtempgen}) decreases with growing parameter ratios $r_n$, this even leads to an AFL behaviour where the chemical potential has left the band as can be seen in Fig.~\ref{fig:phasediagram}.

	Reviewing the results of this section characteristic (breakdown) temperatures were found in Eqs.~(\ref{eq:fermitemp}), (\ref{eq:sommerfeldtemp}), (\ref{eq:boltzmanntemp}), (\ref{eq:degtemp})-(\ref{eq:t0}) which may be ordered as
	\begin{equation}
		T_S < T_{\mathrm{deg}} < T_0 < T_B < T_P \,.
	\end{equation}
	While for large discontinuities $T_F$ lies between the Sommerfeld breakdown temperature $T_S$ and the degeneracy temperature $T_{\mathrm{deg}}$ it becomes larger than $T_{\mathrm{deg}}$ for small ones.
	In contrast to the AFL by Boltzmann approximation, where the temperature has only to be larger then the Boltzmann breakdown temperature $T_B$, the AFL by APLD is centred around the temperature ${\tilde T}_P^{(-)}$ from Eq.~(\ref{eq:expTPlowr}) for vanishing parameter ratios $r_n$ or slightly above for finite values of it, and crucially depend on the influence of the discontinuity.
	This can be seen in the ($r_1$,$y$)-phase diagram as found in Fig.~\ref{fig:phasediagram}.
	Of course in a more realistic density of states the presence of an upper band can destroy these phases if the chemical potential becomes too close to it\footnote{A corresponding Taylor series of an upper band separated by a gap $\Delta$ would lead in Eq.~(\ref{eq:exact}) similar to a lower bandwidth to polylogarithm functions with argument $\nu\mathrm{e}^{\beta \Delta}$.
	Therefore the terms of the approximations would be corrected only by exponentially suppressed terms which are negligible if the chemical potential remains smaller than the size of the gap.}.
	Therefore a large enough gap is necessary, too.

\section{Application}\label{sec:application}

	In this section the formulas obtained above are applied to doped CuCrO${}_{\mathrm{2}}$ and doped CuRhO${}_{\mathrm{2}}$.
	Since obtaining the chemical potential experimentally is a challenging task, the temperature dependence of the thermopower should be used to determine this potential within the TICR framework and, with the results of the last section, an effective density of states.
	However, fitting an infinite number of parameters as appearing in its Taylor series, Eq.~(\ref{eq:Taylorrho}), is obviously not practicable.
	Instead, only the first two terms of the Taylor series should be considered.
	In addition, the replacement of chromium by magnesium as well as the substitution of rhodium later leads to a hole-doping in the $t_{2g}$ band.
	For this kind of doping the upper band edge is important in contrast to the lower one which was described in Eq.~(\ref{eq:Taylorrho}) for electron-like doping.
	In order to apply the previously discussed framework, the signs of all energies should therefore be inverted.
	This leads to the density of states to be given by
	\begin{equation}\label{eq:rho}
		\rho(\varepsilon) = (\rho_0 - \rho_1 \varepsilon) \cdot \Theta(-\varepsilon)\,.
	\end{equation}
	Furthermore, it will turn out, that the effective charge carrier density is accessible, too.

  \subsection{The TICR framework}\label{sec:ticr}

	The thermopower can be studied within the approach of the temperature independent correlation functions ratio approximation (TICR) \cite{Maignan092,Fresard02}.
	This approximation is derived from the Kubo formula where the thermopower is given by \cite{Chaikin76,Luttinger64}
	\begin{equation}
		S = \frac{1}{q_eT} \left( \frac{\langle j_E j_n \rangle}{\langle j_n j_n \rangle} - \mu \right)\,,\label{eq:kubo}
	\end{equation}
	where $q_e$ is the negative charge of the electron and the particle (energy) current operator is denoted by $j_n$ ($j_E$).
	If a shift of the single-particle energy is considered weight can be shifted between both terms on the right side of this formula.
	In particular a shift $E_0$ exists, where the transformed correlation function between $j_n$ and $j_E$ vanishes
	\begin{equation}
		\langle j_E' j_n' \rangle = \langle j_E j_n \rangle + E_0 \langle j_n j_n \rangle = 0\, .
	\end{equation}
	Therefore the thermopower is only given by the transformed chemical potential \mbox{$\mu' = \mu - E_0$}:
	\begin{equation}\label{eq:ticr}
		S = \frac{1}{q_eT} \big( E_0 - \mu \big)\,.
	\end{equation}
	Comparing this result to Eq.~(\ref{eq:kubo}) determines the shift
	\begin{equation}
		E_0 = \frac{\langle j_E j_n \rangle}{\langle j_nj_n \rangle}\,,
	\end{equation}
	which would be temperature dependent in general.
	Thus although Eq.~(\ref{eq:ticr}) describes the transport function $S$ in terms of the equilibrium property $\mu'$, for the full temperature dependence transport theory is still necessary to determine the shift $E_0$.

	If the transport through a single level is considered, this shift can be determined in case the coupling to ballistic leads is weak and temperature independent.
	If such a level is in resonance, i.e. at the same energy as the chemical potentials of both leads, a temperature gradient would not lead to a current $I$ due to the symmetry of the model.
	This means that the thermopower would vanish.
	Thus in the following will be considered the opposite situation where the level would be far away from the chemical potentials.
	In this limit, the Lorentzian transmission $\mathcal{T}(\varepsilon)$ for the non-resonant level model might be expanded in terms of the broadening $\Gamma$ of the level
	\begin{equation}
		\mathcal{T}(\varepsilon) = \Gamma \,\delta(\varepsilon - \varepsilon_0) + \mathcal{O}(\Gamma^2)\,,
	\end{equation}
	where $\varepsilon_0$ denotes the energy of the level and the normalisation at zero temperature was taken into account.
	The Landauer formula\cite{Buttiker85,Hu85} for the tunneling current $I$ gives for such a transmission in the spinless case
	\begin{align}
		I &= \frac{q_e}{\pi \hbar} \int \mathcal{T}(\varepsilon) \big[f_L(\varepsilon-\mu_L) - f_R(\varepsilon-\mu_R)\big]\,\mathrm{d} \varepsilon\\[0.2em]
		  &= \frac{q_e\Gamma}{\pi \hbar} \big[f_L(\varepsilon_0-\mu_L) - f_R(\varepsilon_0-\mu_R)\big]\,,
	\end{align}
	where $\hbar$ is the reduced Planck constant and the Fermi function $f$ as well as the chemical potential $\mu$ in the left and right lead where distinguished by the indices $L$ and $R$ as will be later on for the temperature as well.
	For the condition that no current flows $I=0$, as assumed in the definition of the thermopower, the Fermi functions and therefore their arguments have to be equal
	\begin{equation}
		(\varepsilon_0-\mu_R)\,k_{\mathrm{B}} T_L = k_{\mathrm{B}} T_R \,(\varepsilon_0-\mu_L)\,.
	\end{equation}
	In expressing therein the quantities of the left and right environment by their small differences $q_e\Delta V$ and $\Delta T$ as well as the averaged temperature $T$ and chemical potential $\mu$, the thermopower is obtained as
	\begin{equation}
		S = \frac{\Delta V}{\Delta T} = \frac{1}{q_e T}\big( \varepsilon_0 - \mu \big)\,.
	\end{equation}
	Therefore the $E_0$ function assumes the constant position of the considered level $\varepsilon_0$.
	Furthermore, the same relation can be derived when the transmission is governed by thermal processes\cite{Segal05}.
	In addition, it can be shown that this shift is not important at very high temperatures\cite{Chaikin76}.

	Motivated by these results the basic assumption of TICR treats the shift $E_0$ independent of temperature in the considered temperature region.
	Especially when the chemical potential is inside the band gap where the structure of the bands should not be important, the situation is similar to the case of the non-resonant level model.
	Therefore the approximation should be valid in this case.
	The TICR can thus be viewed as a generalization of a purely stochastic approach \cite{Chaikin76,Koshibae00,Koshibae01} where the first order corrections in the broadening $\Gamma$ is taken into account.
	However, if the thermopower for a usual three-dimensional dispersion relation is calculated in Fermi liquid theory\cite{Behnia04}, it deviates from the result given by TICR for the chemical potential following from such a dispersion relation.
	This means the TICR will break down at low temperatures for a metal when the chemical potential is inside the band.

	In contrast to earlier studies where the shift $E_0$ was set to the Fermi energy \cite{Peterson07,Marsh96} or omitted \cite{Chaikin76,Koshibae00,Koshibae01} it has proven useful in recent studies \cite{Maignan092,Fresard02,Miclau07} to keep it as a free parameter, especially at large temperatures.
	Furthermore, this constant may account for some of the contributions due to scattering processes caused by disorder which might be important at these temperatures.
	Additionally, it preserves the freedom of choosing an arbitrary zero-point of the single-particle energies.
	In the following investigation the stated values of the constant will be with respect to the choice made in Eq.~(\ref{eq:rho}) while the chemical potential is given by the expressions determined in the last section.
	If, as found there, a chemical potential of an apparent Fermi liquid (AFL) is assumed,
	\begin{equation}
		\mu = {\tilde \varepsilon}_F - q_eS_0 T + \varpi T^2\,,
	\end{equation}
	the thermopower will be $T$-linear with a hyperbolic and a constant offset:
	\begin{equation}
		S(T) = \frac{E_0 - {\tilde \varepsilon}_F}{q_eT} + S_0 + \frac{\varpi}{|q_e|} T\,.\label{eq:SaflTICR}
	\end{equation}
	It therefore can describe the observed linear thermopower with an extrapolated offset $S_0$ as shown in Fig.~\ref{fig:AFL} if the hyperbolic behaviour can be neglected due to high temperatures.

	From the expression~(\ref{eq:SaflTICR}) it is easily seen how to determine the linear and quadratic coefficient $S_0$, $\varpi$ of the chemical potential out of the temperature dependence of the thermopower.
	However, the TICR parameter $E_0$ makes it difficult to determine the constant offset.
	While at very large temperatures one can try to neglect the $E_0$ constant,
	\begin{equation}
		|E_0| \ll |{\tilde \varepsilon}_F|\,,\label{eq:largere0}
	\end{equation}
	at finite temperature the study of the quantity
	\begin{equation}
		\frac{T\,S(T) - {\bar T}S({\bar T})}{T - {\bar T}}\,,
	\end{equation}
	where ${\bar T}$ denotes a reference temperature, might be useful since it does not depend on the TICR constant $E_0$ and should be an exact linear function with respect to temperature.
	The experimentally accessible parameters in Eq.~(\ref{eq:SaflTICR}) allows to obtain the parameters of the effective density of states introduced in Eq.~(\ref{eq:rho}) and renormalised in Eq.~(\ref{eq:yrnu}).
	However, their values will depend on the different approximation schemes discussed in Sect.~\ref{sec:theory} and therefore labeled below by an index.
	Furthermore, assuming common parameters of the effective density of states for all samples of an experimental series, e.g. for several manganites \cite{Fresard02}, the exact effective charge carrier density can be obtained in relation to a reference sample.
	Therefore this technique resembles an alternative treatment to identify the doping values of such a series complementary to other experimental based techniques like EDS.

	Combining the TICR expression~(\ref{eq:ticr}) with the result of the APLD given in Eq.~(\ref{eq:muHeikes}) for only a discontinuity in the density of states in the limit of large temperatures yields the famous Heikes formula in the atomic non-degenerate limit $\rho_0W=1$, or modified version of it like derived in \cite{Chaikin76} with $g=\rho_0W$ as the  degenerate factor.
	Furthermore, the combination of the TICR with the result of Boltzmann approximation in Eq.~(\ref{eq:boltzmannnoapprox}) leads to the observation, that the thermopower still resembles the doping dependence of this formula at low doping $S \sim -\ln \mathrm{x}$.
	The prefactor can also be related to the findings reported for the thermopower in the atomic limit of degenerate energy levels.
	For a lattice consisting of atoms with $g$ fully occupied degenerate states a doping with atoms missing one fermion would lead to a $g$-fold degeneracy of the ground-state of the latter atoms.
	As discussed in \cite{Koshibae01} this would lead to a thermopower of
	\begin{equation}
		S_{\mathrm{atomic}} = \frac{k_{\mathrm{B}}}{q_e} \ln \left( g\,\frac{1-\mathrm{x}}{\mathrm{x}} \right)
	\end{equation}
	in the high temperature limit.
	When fermions delocalise the distribution in the density of states broadens.
	However, since the total number of states remains fixed a discontinuity in a density of states like Eq.~(\ref{eq:rho}) of
	\begin{equation}
		\rho_0 = \frac{g}{W}
	\end{equation}
	is necessary if the slope therein is not considered.
	According to Eqs.~(\ref{eq:boltzmann}) and~(\ref{eq:SaflTICR}) the thermopower will be given by
	\begin{equation}
		S = \frac{k_{\mathrm{B}}}{q_e} \ln \left(\frac{k_{\mathrm{B}} {\tilde T}}{W}g \,\frac{1}{\mathrm{x}}\right) + \frac{k_{\mathrm{B}}}{q_e} \frac{1}{2{\tilde T}} T\,,
	\end{equation}
	if the hyperbolic terms is omitted due to high temperatures.
	While for large expansion temperatures ${\tilde T}$ the linear term will be negligible the degeneracy factor of the constant term is renormalised compared to the atomic limit.
	Therefore in considering a broad band only the thermally accessible region of the band enters in the thermopower.

  \subsection{CuCr\texorpdfstring{${}_{1-\mathrm{x}}$}{1-x}Mg\texorpdfstring{${}_{\mathrm{x}}$}{x}O\texorpdfstring{${}_{\mathrm{2}}$}{2}}\label{sec:cucro2}

	Carrying out the analysis for the strongly correlated delafossite CuCr${}_{1-\mathrm{x}}$Mg${}_{\mathrm{x}}$O${}_2$ gives already reasonable parameters of the density of states using Boltzmann approximation for large parameter ratio, in contrast to Sommerfeld expansion.
	Furthermore, the results coincide with the one obtained by the APLD if a larger expansion temperature $T_0 \rightarrow \mu\,(2k_{\mathrm{B}})^{-1}$ is used indicating a very low temperature scale.
	This requires to take the first order in this approximation proportional to $d_1(\nu_0)$ in Eq.~(\ref{eq:diff}) into account, too.

	\begin{figure}[t!]
		\centering
		\includegraphics[width=\columnwidth]{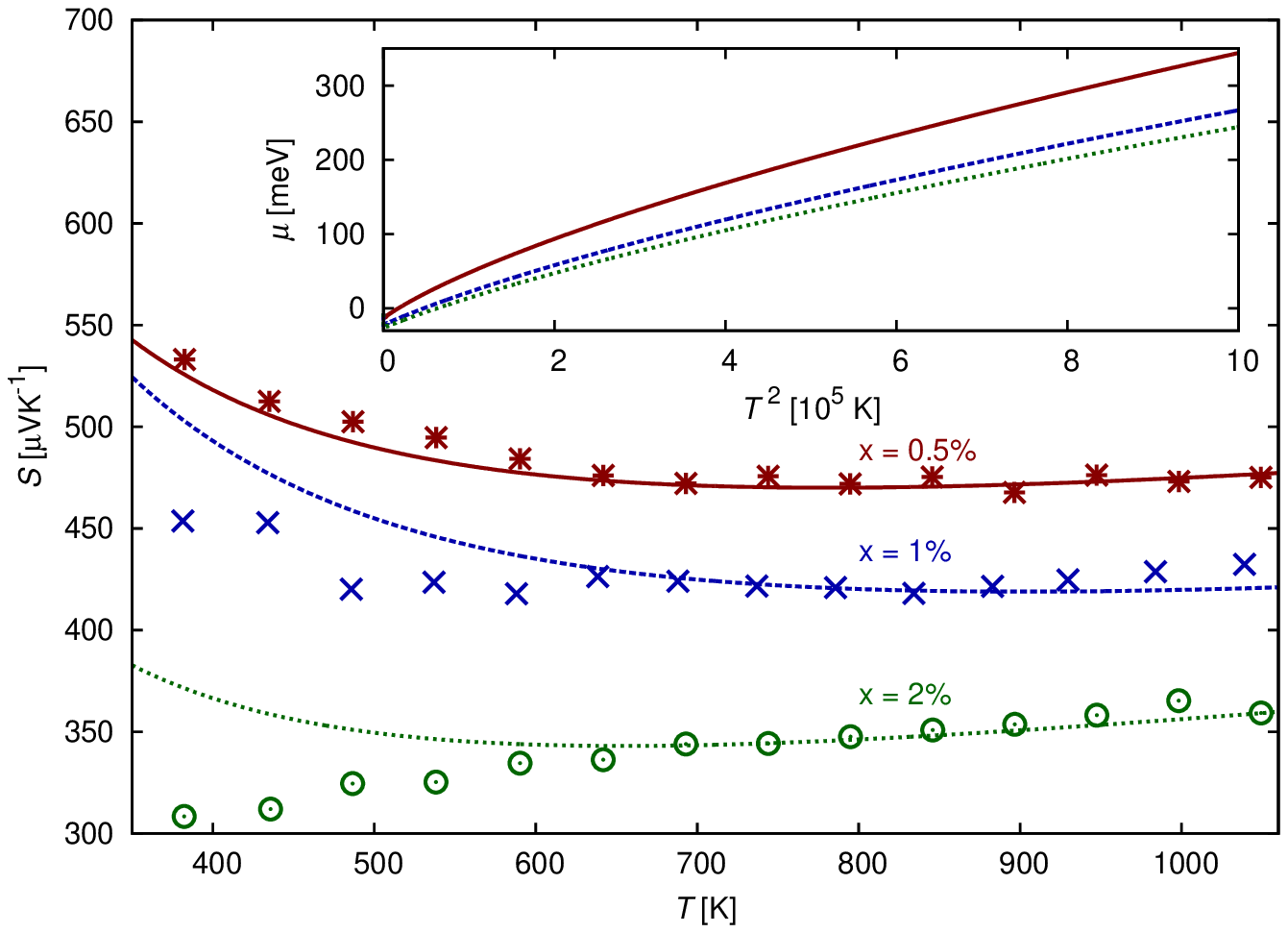}
		\caption{
			Comparison of the thermopower of CuCr${}_{1-\mathrm{x}}$Mg${}_{\mathrm{x}}$O${}_2$ given by experimental data for nominal doping $\mathrm{x}=0.5\%$ (stars), $\mathrm{x}=1\%$ (crosses) and $\mathrm{x}=2\%$ (circles) to the one obtained from the parameters of the density of states Eq.~(\ref{eq53}) for values of the effective charge carrier density of $0.35\%$ (solid line, with $E_0=134\,\mathrm{meV}$), $0.85\%$ (dashed line, with $E_0=137\,\mathrm{meV}$), and $1.1\%$ (dotted line, with $E_0=105\,\mathrm{meV}$).
			\mbox{Inset: Temperature} dependence of the chemical potentials for the corresponding parameter values.
			A quadratic behaviour can be observed at very low and very high temperatures.
		}
		\label{fig:SCuCrO2}
	\end{figure}
	While the results of the TICR parameter $E_0=100\,\mathrm{meV}$ are close to the ones reported for doped manganites, but with opposite signs since they are electron-doped \cite{Fresard02,Miclau07}, the slope $\rho_1$ seems to dominate the effective density of states for this material.
	Although this is in agreement with a previous first principle study \cite{Maignan091}, it challenges the determination of the small discontinuity $\rho_0$.
	Furthermore, the results exhibit a strong increase of both parameters for the highest doped samples which falls in the range where the formation of a secondary spinel or CuO phase was previously reported \cite{Maignan091,Guilmeau11}.
	They are therefore omitted in the following investigation.
	The other samples show still results of the parameters of the effective density of states depending on the doping in contrast to the parameter ratio $r_1=200\,(\mathrm{eV})^{-1}$.
	This encourages adjusting the doping to find the effective charge carrier density, as was done for several other manganites \cite{Fresard02}.
	For this material this problem was experimentally addressed in a previous publication \cite{Guilmeau11} leading to quite different doping values by statistical electron diffraction spectroscopy (EDS).
	If the most reliable value of the reported data $\mathrm{x} = 1.1\%$ is used for the nominal one of $\mathrm{x} = 2\%$ the parameters of the density of states,
	\begin{equation}\label{eq53}
		\rho_{0,B} = 0.121\,\mathrm{(eV)^{-1}}\qquad \mathrm{and} \qquad \rho_{1,B} = -24\,\mathrm{(eV)^{-2}}\,,
	\end{equation}
	coincide for all considered samples for effective charge carrier density of
	\begin{equation}
		\mathrm{x} = 0.35\%,\ 0.85\%,\ 1.1\%\,,
	\end{equation}
	which are within the error margins of the one suggested by the EDS measurements.
	Due to this renormalisation the $E_0$ parameter are only slightly modified during the minimization process.

	The calculation of the thermopower using the exact numerical solution leads to good agreement between theory and experiment as can be seen in Fig.~\ref{fig:SCuCrO2}, if the TICR constant $E_0$ is adjusted to the modified doping values by an additional fit.
	Additionally, the hyperbolic offset caused by the TICR parameter seem to describe the increase at low temperature and doping, while the agreement is lost below $T \sim 600\,\mathrm{K}$ for intermediate doping values.
	This might be due to a breakdown of TICR, but surprisingly coincides with the Boltzmann breakdown temperatures for these samples \mbox{$T_B=356\,\mathrm{K},\ 497\,\mathrm{K}$ and $608\,\mathrm{K}$}.
	These values confirm the placement of the observed behaviour in the AFL by Boltzmann approximation, too.

  \subsection{CuRh\texorpdfstring{${}_{1-\mathrm{x}}$}{1-x}Mg\texorpdfstring{${}_{\mathrm{x}}$}{x}O\texorpdfstring{${}_{\mathrm{2}}$}{2}}\label{sec:curho2}

\begin{figure}[t!]
		\centering
		\includegraphics[width=\columnwidth]{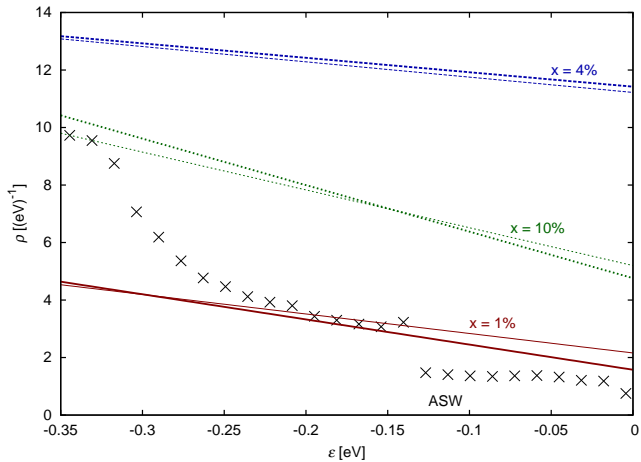}
		\caption{
			Density of states of CuRh${}_{1-\mathrm{x}}$Mg${}_{\mathrm{x}}$O${}_{\mathrm{2}}$ as given by the augmented spherical wave 2 method (crosses) and their shape as determined by a fit to the exact form Eq.~(\ref{eq:apldnoapprox}) (thick) and quadratic expansion Eqs.~(\ref{eq:apldrenormalizationstart})-(\ref{eq:apldrenormalizationend}) (thin) in APLD for different values of the doping of \mbox{$\mathrm{x} = 1\%$} (solid), $4\%$ (dashed) and $10\%$ (dotted).
		}
		\label{fig:dos}
	\end{figure}

\begin{figure}[t!]
		\centering
		\includegraphics[width=\columnwidth]{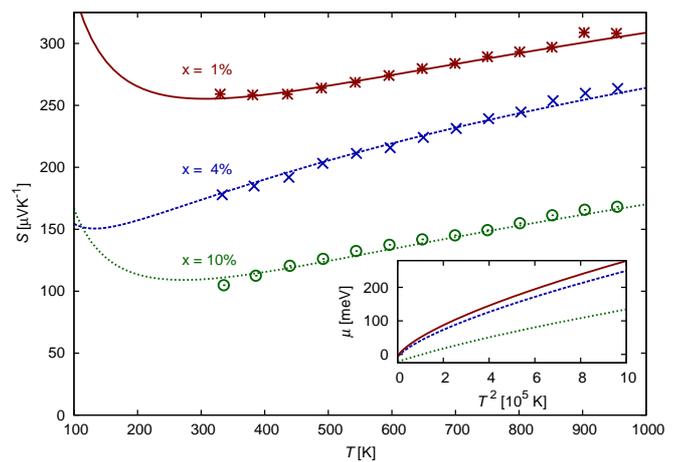}
		\caption{
			Comparison of the thermopower of the delafossite CuRh${}_{1-\mathrm{x}}$Mg${}_{\mathrm{x}}$O${}_{\mathrm{2}}$ given by experimental data for nominal doping $\mathrm{x}=1\%$ (stars), $\mathrm{x}=4\%$ (crosses) and $\mathrm{x}=10\%$ (circles) to the one obtained from the parameters of the density of states Eq.~(\ref{eq:resultapld}) for values of the effective charge carrier density of $2\%$ (solid line, with $E_0=30\,\mathrm{meV}$), $3\%$ (dashed line, with $E_0=15\,\mathrm{meV}$) and $10\%$ (dotted line, with $E_0=36\,\mathrm{meV}$).
			Inset: Temperature dependence of the chemical potentials for the corresponding parameter values.
			Two quadratic regions can be identified, one described by Fermi liquid theory at low temperature and one in the framework of an AFL at higher temperatures similar to the thermopower and conductance shown in Fig.~\ref{fig:AFL}.
		}
		\label{fig:S}
	\end{figure}

	The application of the presented technique to the measurements of the doped band insulator CuRhO${}_2$\cite{Maignan092} using established approximations does not give cohesive results.
	In comparing the results stated in Tab.~\ref{tab:expvstheo} and produced by a fit using the formula of the APLD with the expansion temperature given in Eq.~(\ref{eq:t0}), the values of the parameter of the TICR $E_0\sim 30\,\mathrm{meV}$ are in agreement for the lowest and highest doped samples and fall in a region previously reported \cite{Maignan092,Fresard02}, too.
	Furthermore, the results from the expansion for small parameter ratio obtained in Eq.~(\ref{eq:apldsmallr}) give values similar to the ones obtained from the quadratic expansion and from fitting the full solution stated in Eq.~(\ref{eq:apldnoapprox}).
	Again the found parameter ratios are close, while the values of the effective doping vary.
	Therefore the doping can be adjusted where the results of the parameters of the density of state coincide for values of
\begin{table}
	\centering
	\begin{tabular}{@{}l@{}l@{\,}|@{\,}l@{\,}l@{~~}l@{\,}|@{\,}l@{~~}l@{~~}l@{\,}|@{\,}l@{~~}l@{~~}l@{}}\hline
		&& \multicolumn{3}{@{}c@{~}|@{~}}{$\mathrm{x} = 1\%$} & \multicolumn{3}{c@{~}|@{~}}{$\mathrm{x} = 4\%$} & \multicolumn{3}{c@{}}{$\mathrm{x} = 10\%$}\\
		&& $\mu_{P}$ & $\mu^{(q)}_{P}$ & $\mu^{(-)}_{P}$  &  $\mu_{P}$ & $\mu^{(q)}_{P}$ & $\mu^{(-)}_{P}$  &  $\mu_{P}$ & $\mu^{(q)}_{P}$ & $\mu^{(-)}_{P}$ \\ \hline
		$r_1$&[(eV)${}^{\mathrm{-1}}$] & 3.5 & 3.1 & -    &  3.4 & 3.4 & 3.4  &  3.4 & 2.5 & - \\
		$y$&[meV]                      & 4.8 & 4.6 & 2.4  &  6.0 & 6.0 & 3.1  &  21  & 19  & 13\\
		$E_0$&[meV]                    & 33  & 31  & 6.1  &  11  & 11  & -14  &  36  & 32  & 17\\\hline
	\end{tabular}
	\caption{
		Stationary points of the fitting routine when adjusting the parameters in the formulas of the APLD in a region from $600\,\mathrm{K}$ to $1000\,\mathrm{K}$ to the experimental data of CuRh${}_{1-\mathrm{x}}$Mg${}_{\mathrm{x}}$O${}_{\mathrm{2}}$.
	}
	\label{tab:expvstheo}
\end{table}
	\begin{equation}
		\mathrm{x} = 2.3\%,\ 2.8\%,\ 10\%\,.
	\end{equation}
	With these values of the doping all samples give therefore rise to the same parameters of the density of states of
	\begin{equation}
		\rho_{0,P}=4.7\,\mathrm{(eV)}^{-1} \qquad \mathrm{and} \qquad\rho_{1,P}=-16\,\mathrm{(eV)}^{-2}\,.\label{eq:resultapld}
	\end{equation}
	As can be seen in Fig.~\ref{fig:dos}, the results of the APLD are close to the ones obtained by first principle studies.
	The extracted parameters places the largest doped sample of this material in the phase diagram shown in Fig.~\ref{fig:phasediagram} precisely in the narrow area (intermediate parameter ratio $r_1$) of the validity range of the AFL by APLD, while the lowest doped samples are closer to the region described by Boltzmann approximation.
	This is further supported by the fact that the Boltzmann breakdown temperature for the largest doped sample \mbox{$T_B(\mathrm{x}=10\%) = 1316\,\mathrm{K}$} is larger than the one considered while the expansion temperature of the APLD \mbox{$T_0(\mathrm{x}=10\%) = 653\,\mathrm{K}$} lies within the region where the linear behaviour is observed.
	Therefore the observed AFL can be understood as the remains of a larger manifestation for larger discontinuity.
	Compared to doped CuCrO${}_2$ this material exhibits therefore a discontinuity enhanced thermopower although larger doping values were considered.
	Furthermore, the recalculation of the thermopower using the exact solution shows extraordinary good agreement as can be seen in Fig.~\ref{fig:S}.
	At low temperature the hyperbolic offset $E_0$ of the thermopower dominates the theoretical behaviour for this material, too.
	Nevertheless, since the data does not show a strong upturn this rather indicate the breakdown of the TICR.
	However, a slight upturn can be found for the lowest doped sample which compared to the intermediate one has a larger $E_0$ constant.
	Since interaction effects are expected to be negligible due to nearly filled bands this might arise from scattering processes caused by a larger disorder in this sample which could explain the insulating behaviour at very low temperatures \cite{Maignan092}, too.

\begin{figure}[t!]
		\centering
		\includegraphics[width=\columnwidth]{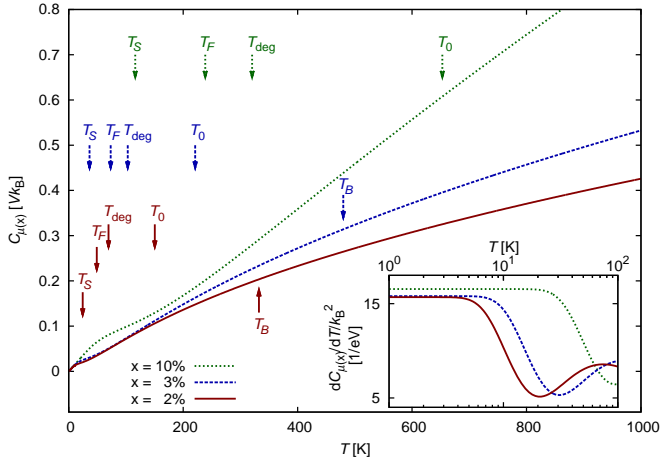}
		\caption{
			Specific heat of CuRh${}_{1-\mathrm{x}}$Mg${}_{\mathrm{x}}$O${}_{\mathrm{2}}$ and its derivative with respect to the temperature (inset) given by the calculation with the parameters of the density of states from Eq.~(\ref{eq:resultapld}) for values of the doping $\mathrm{x}=2\%$ (solid), $3\%$ (dashed) and $10\%$ (dotted).
			The arrows indicate the position of the characteristic (breakdown) temperatures for the different doping values.
			The curves show a transition between a Fermi liquid and an apparent Fermi liquid indicated by an inflexion point.
		}
		\label{fig:cv}
	\end{figure}

	From the parameters of the density of state further quantities like the specific heat can be determined which is given in Fig.~\ref{fig:cv}.
	Since the determination of the specific heat is not dependent on the TICR parameter a Fermi liquid behaviour can be observed at very low temperature.
	Nevertheless, for low doping the effective mass gets renormalised in comparison to calculations in which the discontinuity is neglected.
	This phase breaks down at a temperature which decreases with decreasing doping, in agreement with the expected insulating behaviour for zero doping.
	However, the breakdown temperatures are obviously not well described by the Fermi temperature $T_F$, but seem to occur at temperatures somewhat below the analytically introduced smaller breakdown temperature $T_S$.
	In the vicinity of this temperature a peak in the derivative of the specific heat is found, as shown in the inset of Fig.~\ref{fig:cv}.
	It should be observable in experiments, in particular for low doping.
	Above the transition area, around the degeneracy temperature $T_{\mathrm{deg}}$, a linear region emerges which extends to $T_0$, where the chemical potential has left the band, and for large doping even beyond.
	Note that the difference between the apparent effective mass and the bare one exceeds the one usually expected from interaction effects obtained using the slave-boson saddle-point approximation applied to a nearly filled band \cite{Fresard97}.
	At higher temperatures for the two lower doped samples, the slope gradually reaches a slightly smaller value, forming another linear region above the breakdown temperature of Boltzmann approximation $T_B$.
	This region is described by the AFL in Boltzmann approximation.
	Under an increase of the doping, the APLD replaces Boltzmann approximation as the valid approximation at intermediate temperature.
	In this way the transition from the AFL in Boltzmann approximation to the one given by the APLD is visible.
	This can be seen in Fig.~\ref{fig:phasediagram}, too, where an expansion around \mbox{${\tilde T}=800\,\mathrm{K}$} was considered and the difference of the chemical potential at a temperature \mbox{$T = 400\,\mathrm{K}$} was studied.
	The transition occurs therein when moving upwards, respectively increasing the renormalised doping $y$, along the line for \mbox{$r_1=3.4\,\mathrm{(eV)^{-1}}$}.
	Since the transition temperatures $T_0$ for the different doping values are always lower than the expansion temperature ${\tilde T}$ the APLD as described by Eq.~(\ref{eq:apldnoapprox}) remains valid.

\section{Conclusion}\label{sec:conclusion}

	This investigation of the origin of an AFL behaviour started from a density of states of non-interacting fer\-mi\-ons written as a Taylor series around a single discontinuous band edge.
	With this simplification Sommerfeld expansion and Boltzmann approximation could be formulated as expansion of the polylogarithm.
	While the explanation of Fermi liquid like behaviours at large temperature within the Fermi liquid theory would require large temperature scales, it can be expected from the latter approximation for very low temperature scales, too, e.g. in the chemical potentials shown in the insets of Figs. 5 and 7.
	However, the behaviour of this approximation presents distinctive differences like an offset in the thermopower within the TICR framework.
	Furthermore, its doping dependence is similar to the one of Heikes behaviour but does not describe such a temperature independent behaviour.
	In addition, it states that the degeneracy factor in modified Heikes formulas has to be modified for broad bands by the fraction which is thermally accessible.

	An analytic anomaly, like the discontinuous band edge, further influences physical properties.
	Indeed it causes the Fermi liquid to break down before the temperature reaches the Fermi temperature.
	The missing region was made accessible by the APLD.
	The lowest order of this approximation treats the system at a given temperature with an averaged density of states and a renormalised doping.
	While it incorporates Boltzmann approximation, it becomes exact if the density of states can be described only by a discontinuity.
	In this case the large temperature limit results in the modified Heikes formula and therefore describes the temperature independent thermopower as given by Heikes formula correctly.
	Furthermore, an additional AFL region could be identified by the APLD at a particular temperature.
	This phase strongly depends on the presence of a noticeable discontinuity.
	Similar to the lower tail of the Fermi function in Boltzmann approximation, its upper tail combined with the missing states beyond the discontinuous band edge will then lead to an AFL behaviour, but at lower temperatures than the former approximation.

	By the application of this framework on the doped delafossites CuCrO${}_2$ and CuRhO${}_2$, one material could be identified which falls in the Boltzmann region and one which exhibits an AFL behaviour caused by a large discontinuity.
	The data of the thermopower of these materials show very good agreement to the presented theory.
	In addition, the recently addressed formation of a spinel or CuO phase in the former material was indicated by the change of the parameters of the effective density of states.
	Furthermore, the determination of the effective charge density could be performed within the phenomenological treatment the results of which were in good agreement with previous EDS measurements.
	From the latter material the effective doping values were determined, too, and a much larger discontinuity was extracted which coincide with ASW2 calculation.
	Thereby it was seen that although larger doping values were considered than for the former material the thermopower was additionally enhanced by the discontinuity.
	Additionally, this anomaly would result in a peak in the derivative of the electronic part of the specific heat of this material which might be experimentally tested.

\section*{Acknowledgement}

	We would like to gratefully thank V.~Eyert for useful discussions and for sharing his results of the density of states obtained by the augmented spherical wave~2 method,
	E.~Guilmeau and A.~Maignan for having provided us with the thermopower data of CuCr${}_{1-\mathrm{x}}$Mg${}_{\mathrm{x}}$O${}_{\mathrm{2}}$ and CuRh${}_{1-\mathrm{x}}$Mg${}_{\mathrm{x}}$O${}_{\mathrm{2}}$.
	This work was supported by the ANR through NEWTOM-08-BLAN-0005-01, by the R\'egion Basse-Normandie, by the Minist\`ere de la Recherche, by the Deutsche For\-schungs\-gemeinschaft through the Research Unit 960 ``Quantum Phase Transitions'' and by the FGU.
\enlargethispage{0.7em}

\end{document}